\documentclass[twocolumn,aps,prb,floatfix]{revtex4}
\usepackage{graphicx,bm,times,amsmath}
\allowdisplaybreaks
\newcommand{\I} {\mathrm{I}}
\newcommand{\II}{\mathrm{I\hspace{-.1em}I}}
\begin{document}

\title{
Intramolecular charge ordering in
the multi molecular orbital  
system  (TTM-TTP)I$_3$ 
}

\author{
Marie-Laure Bonnet,$^1$ 
Vincent Robert,$^1$ 
Masahisa Tsuchiizu,$^2$
Yukiko Omori,$^2$ and 
Yoshikazu Suzumura$^2$}
\affiliation{
${}^1$Laboratoire de Chimie,
Ecole Normale Sup\'erieure de Lyon, 
CNRS, 46 all\'ee d'Italie, F-69364 Lyon, France
\\
${}^2$
Department of Physics, Nagoya University, Nagoya 464-8602, Japan
}

\date{June 8, 2010: J.\ Chem.\ Phys.\ \textbf{132}, 214705 (2010)}

\begin{abstract}
  Starting from the structure of the (TTM-TTP)I$_3$ molecular-based
  material, we examine the characteristics of frontier molecular
  orbitals using \textit{ab initio} (CASSCF/CASPT2) configurations
  interaction calculations.  It is shown that the singly-occupied and
  second-highest-occupied molecular orbitals are close to each other,
  i.e., this compound should be regarded as a two-orbital
  system.  By dividing virtually the [TTM-TTP] molecule into three
  fragments, an effective model is constructed to rationalize the origin
  of this picture.  In order to investigate the low-temperature symmetry
  breaking experimentally observed in the crystal, the electronic
  distribution in a pair of [TTM-TTP] molecules is analyzed from CASPT2
  calculations.  Our inspection supports and explains the speculated
  \textit{intramolecular} charge ordering which is likely to give rise
  to low-energy magnetic properties.
\end{abstract}

\maketitle

\section{Introduction}
The characterization of molecule-based electronic
conductors is one of the central issue in the research of molecular
crystal systems.  Typical materials are TTF-TCNQ and (TMTSF)$_2$PF$_6$,
where TTF, TCNQ, and TMTSF stand for 
tetrathiafulvalene, tetracyanoquinodimethane, and 
tetramethyltetraselenafulvalene,
 respectively.
In (TMTSF)$_2$PF$_6$,
the first superconductivity behavior in molecular
solids has been reported.  \cite{Jerome1982} Recent hot topics in the
research of molecular crystals is the realization of charge ordering
phenomena.  \cite{Seo2004} Since the pioneer work of
 Su \textit{et al.}\cite{ssh}
in polyacetylene,  the charge trapping phenomenon has been
much studied in Peierls transition issues.  \cite{peierls} In that
sense, quasi-one-dimensional chains have received much attention from
experimental and theoretical points of
view.\cite{borshch1998a,robert1999a,robert2004a,malrieu2004a}
Theoretical approaches that focus on a single
highest-occupied-molecular orbital (HOMO) or lowest-unoccupied-molecular
orbital (LUMO) have been successful in describing fascinating electronic
ordered phase.  Such a treatment can be justified since in these
conventional systems, the HOMO or LUMO levels are well-separated from
the rest of the MOs spectrum.\cite{Seo2004} 

Nevertheless, new types of
molecular solids have been recently synthesized, for instance, aiming at
the metalization of the single-component molecular solids family
$M$(tmdt)$_2$ ($M =$ Ni, Au),\cite{Tanaka_SCMM2001}
where tmdt stands for trimethylenetetrathiafulvalenedithiolate.
 The molecular
extension of the tmdt system is so large that a description based on a
single molecular orbital (MO) is questionable.  
As a matter of fact, \textit{ab initio}
calculations have been performed
\cite{Rovira,Ishibashi2005,Ishibashi2008} and an effective three-band
Hubbard model has been proposed and succeeded in describing electronic
structures.\cite{Seo2008} In this respect, the quasi-one-dimensional
molecular compound (TTM-TTP)I$_3$, \cite{Mori2004,Mori1994,Mori1997}
where
TTM-TTP=2,5-bis(4,5-bis(methylthio)-1,3-dithiol-2-ylidene)-1,3,4,6-tetrathiapentalene looks like a promising candidate to investigate the
theory limitation of a single-MO approximation.  The presence of an
I$_3^{-}$ counter-anion is suggestive of formally organic
[TTM-TTP]$^{+}$ cations in the crystal structure, i.e.,
unpaired electrons localized within the organic moieties.  The magnetic
susceptibility\cite{Maesato1999,Fujimura1999} and NMR measurements
\cite{Onuki2001_Synth,Onuki2001_JPCS} revealed phase transitions at
finite temperature.  Insulating and non-magnetic behaviors have been
confirmed at low-temperature.  Furthermore, based on Raman-scattering
\cite{Yakushi2003,Swietlik2004a,Swietlik2005} and x-ray measurements,
\cite{Nogami2003} it has been suggested that an asymmetric deformation
of the [TTM-TTP] molecule occurs and charge disproportionation within
the molecule is possible below a transition temperature.  This novel
charge-ordered (CO) state is different from the conventional CO state
\cite{Seo2004} and is called \textit{intramolecular} CO
state.\cite{Swietlik2005} This nontrivial phenomena observed in this
particular compound may not be described using a single-band
description, i.e., the present molecular assembly could give rise
to a phenomenon beyond a concept expected from a knowledge of a single
molecule although the trigger may be hidden in the properties of
multi molecular orbitals. Thus, there is a crucial need for theoretical
description to rationalize the origin of such state.
The purposes of our study are (i) to show that the energy of
singly occupied molecular orbital (SOMO) and that of
second-highest-occupied molecular orbital (HOMO-1) are quasidegenerate,
(ii) to clarify the origin of this quasi-degeneracy using an effective
three-fragment model for the [TTM-TTP] molecule, and (iii) to show explicitly
that the intramolecular charge ordering actually occurs in the neighboring two-molecular
system.  With this goal in mind, multireference wave function-based \textit{ab
  initio} calculations are performed to rationalize the electronic
distribution in the (TTM-TTP)I$_3$ material.

\section{MO{\scriptsize S} of the [TTM-TTP]$^{+}$ ion}

\begin{figure}[t]
\includegraphics[width=5cm]{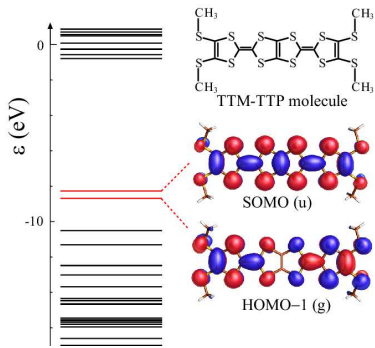}
\caption{
[TTM-TTP] molecule (right-top) and  
energy levels of the isolated [TTM-TTP]$^+$ ion
 from ROHF \textit{ab initio} SCF calculations.
The SOMO (u) and HOMO-1 (g)
  are drawn
where the red and blue colors denote the sign of orbital.
}
\label{fig:mo-level}
\end{figure}

The [TTM-TTP] organic molecule C$_{14}$S$_{12}$H$_{12}$ is shown in Fig.\
\ref{fig:mo-level}.  The atomic coordinates are read from the (TTM-TTP)I$_3$
298~K crystal structure. \cite{Mori1994}
In order to assess the packing influence in the electron
trapping phenomenon, we did not perform any geometry optimization.
Based on this structure, we performed correlated \textit{ab initio}
calculations. This type of approach is very insightful since important
information is accessible through a reading of the wave function. In
particular, complete active space self-consistent field (CASSCF) calculations
have turned out to be very efficient to unravel intriguing electronic
distribution in organic radical-based materials. \cite{kepenekian2009,
  rota2008, messaoudi2006, rota2010} 
On top of the CASSCF wave functions, second-order
perturbation theory calculations (CASPT2) were performed using an
imaginary shift of 0.3 a.u.\ and an 
ionization potential-electronic affinity (IPEA)
shift of 0.0 a.u.
The IPEA shift aims at correcting the energy differences calculations
between states holding different open shells. Since we are dealing
with spin states, it would be irrelevant to turn on this parameter.
\cite{Ghigo,Kepenekian2009}
 This procedure allows one to
incorporate the important dynamical correlation effects 
to reach a high level of accuracy.
All our \textit{ab initio} calculations were
performed using the MOLCAS package \cite{MOLCAS} with all electron basis sets
contractions for the elements S(7s6p1d)/[4s3p1d], C(5s5p1d)/[3s2p1d] and H
(3s)/[1s].  We checked the validity of these particular contractions by
including diffuse and polarization functions which did not lead to any
quantitative changes.

At room-temperature, the [TTM-TTP] molecule exhibits an inversion
center.  Thus, the MOs can be classified as \textit{gerade}
(g) or \textit{ungerade} (u) according to the symmetry point group.  As mentioned
before, the (TTM-TTP)I$_3$ material is a charge-transfer salt,
consisting of [TTM-TTP]$^{+}$ and I$_3^-$ species.
In a simple picture, the HOMO
of the [TTM-TTP]$^{+}$ ion is half-filled, i.e., SOMO.
It has been usually recognized that one may concentrate on this
SOMO, ignoring the rest of the spectrum on the assumption that its energy
is well isolated from those of the other orbitals as compared to the 
bandwidth.\cite{Mori1997}
This is one particular issue we wanted to examine.  
Thus, semiempirical
extended H\"uckel calculations were first
performed.\cite{Mori1984_Huckel,Mori1985_Huckel} 
In the following, the SOMO and HOMO-1 will be
referred to as the u and g orbitals, respectively (see Fig.\
\ref{fig:mo-level}). 
The energy separation
between the g and u valence MOs (see Fig.\ \ref{fig:mo-level}) is
$\approx 0.2$ eV, while the bandwidth of the HOMO is $\approx 1$
eV.\cite{Mori1997} Thus, the effect of the g orbital might not be
negligible since it is likely to participate in the intramolecular CO
phenomenon expected in the (TTM-TTP)I$_3$ system.
In order to clarify the charge distribution, CASSCF calculations were
then carried out allowing the occupation of two MOs by three electrons,
i.e., CAS[3,2]. This method is known to provide very satisfactory charge
distribution as soon as the active space is flexible enough.  The g and
u MOs are then treated on the same footing and both symmetries states
can be examined along these calculations.  The CASSCF energy difference
between the g and u doublets is $\approx 0.5$ eV, confirming the relative
proximity of the frontier orbitals. 
The energy
spectrum was finally calculated using a restricted open-shell SCF
(ROHF) procedure.  
Along these ROHF calculations, three electrons are
likely to occupy the frontier MOs u and g.  
These MOs are likely to be singly occupied or doubly occupied in the 
CASSCF calculations. In order to position the corresponding energies,
we performed a ROHF calculation assuming an average number of electrons
in the g and u MOs, i.e., 1.5 electrons.
The calculated energy levels in the vicinity of the
SOMO are shown in Fig.\ \ref{fig:mo-level}.  From this inspection, the
SOMO has ungerade character, whereas the HOMO-1 is gerade type in agreement
with our extended H\"uckel calculations.  The
respective ROHF energies $\varepsilon_\mathrm{u}=-8.27$ eV and
$\varepsilon_\mathrm{g}=-8.69$ eV, while the energy difference between
the u and g MOs is $\approx 0.4$ eV, a value which is consistent with
our extended H\"uckel estimation.  This combined semiempirical and {\it
ab initio} information upon the constitutive unit [TTM-TTP]$^+$ raises
the relevance of a one-band approach to examine the electronic properties
of (TTM-TTP)I$_3$ crystal.

\section{effective three-fragment model}

\begin{figure}[t]
\includegraphics[width=6cm]{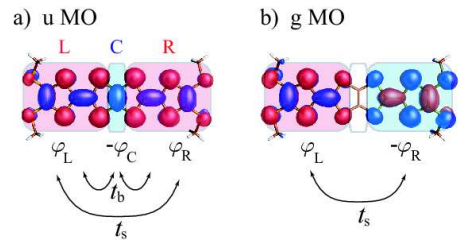}
\caption{
Effective three-fragment model for the u MO
  (a) and the g MO (b).  The three fragments are named as L, R, and C,
  representing the left, right, and center fragments, and their wave functions
  are $\varphi_{\mathrm{L}}$, $\varphi_{\mathrm{R}}$, and
  $\varphi_{\mathrm{C}}$, respectively.  The u MO is given by the
  superposition of $\varphi_{\mathrm{L}}$, $-\varphi_{\mathrm{C}}$, and
  $\varphi_{\mathrm{R}}$, while the g MO is made of $\varphi_{\mathrm{L}}$ and
  $-\varphi_{\mathrm{R}}$.  The mutual interaction between the respective
  fragments are $t_\mathrm{b}$ and $t_{\mathrm{s}}$, representing the
  through-bond and through-space interactions, respectively.  }
\label{fig:frag-model}
\end{figure}

In this section, we analyze the chemical origin of the close-in-energy
character of the g and u MOs.  Let us split the [TTM-TTP] molecule into
3 fragments---L, R, and C---as shown in Fig.\ \ref{fig:frag-model}.  The L
and R parts represent the left and right parts of the [TTM-TTP]
molecule, whereas C corresponds to the ethylene-type bridging moiety.
Based on this fragments picture, the u MO displays ``bonding'' character
while the g MO is ``anti-bonding''.  Therefore, one may wonder why the
bonding MO lies higher in energy than the anti-bonding one.
Part of the answer can be found in the C group orbital u which is likely
to mix in the L-R ``bonding'' orbital (see Fig.\ \ref{fig:frag-model}).
As for the interaction between these fragments, we consider two
types of hopping integrals here, namely, $t_{\mathrm{b}}$ and
$t_{\mathrm{s}}$.  The former accounts for the through-bond interaction,
while the latter represents the through-space interaction.  The energy
levels of the isolated L and R fragments are identical, and set to
$\varepsilon_0$.  On the other hand, the energy level of the isolated C
fragment is much lower than the L and R fragments, and we parametrize
the energy difference as $\Delta \varepsilon$ $(>0)$.  
Using the local  orbitals
$\varphi_{\mathrm{L}}$, $\varphi_{\mathrm{R}}$, and 
$\varphi_{\mathrm{C}}$
shown in Fig.\ \ref{fig:frag-model}, 
the effective Hamiltonian for the 3-fragment model reads
\begin{eqnarray}
\left(
\begin{array}{ccc}
\varepsilon_0     & - t_{\mathrm{s}} & - t_{\mathrm{b}}
\\
 - t_{\mathrm{s}} &  \varepsilon_0   & - t_{\mathrm{b}}
\\
 - t_{\mathrm{b}} &  - t_{\mathrm{b}} &   \varepsilon_0 -\Delta \varepsilon
\end{array}
\right)
\begin{array}{c}
\mbox{\scriptsize L}\phantom{Z}
\\
\mbox{\scriptsize R}\phantom{Z}
\\
\mbox{\scriptsize C}\phantom{Z}
\end{array}
\end{eqnarray}

\begin{figure}[b]
\includegraphics[width=7cm]{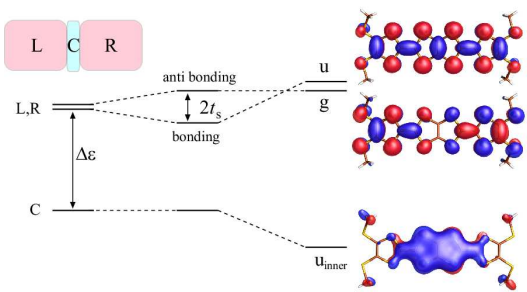}
\caption{ MOs diagram for the effective three-fragment model of the
  [TTM-TTP] molecule.  As the through-space interaction $t_{\mathrm{s}}$ is
  turned on, the L and R local orbitals mix in and result in bonding and
  antibonding MOs (center).  Then, the through-bond interaction
  $t_{\mathrm{b}}$ pushes the bonding MOs higher in energy (right).  The
  identified MOs, u, g, and, u$_{\mathrm{inner}}$, which are obtained from the
  SCF calculations, are also shown.  The respective eigen-energies are given
  by $\varepsilon_\mathrm{u}=-8.27$ eV,
    $\varepsilon_\mathrm{g}=-8.69$ eV, and
    $\varepsilon_\mathrm{u_{inner}}=-19.59$ eV.  
}
\label{fig:energydiagram-mo}
\end{figure}

First, let us neglect the through-bond $t_{\mathrm{b}}$ interaction.  The
through-space interaction between the L and R parts results in bonding and
antibonding MOs whose energy difference is $2t_{\mathrm{s}}$.  This virtual
situation is shown in the center of Fig.\ \ref{fig:energydiagram-mo}.  As soon
as the $t_{\mathrm{b}}$ interaction is turned on, the energy of the bonding MO
is shifted higher, while the anti-bonding MO remain unchanged due to
symmetry constraints.  Assuming that $\Delta \varepsilon$ is much larger
than $|t_{\mathrm{b}}|$, the eigenvalues read
\begin{subequations}
\begin{eqnarray}
\varepsilon_{\mathrm{u}}
&\approx&
\varepsilon_0 - t_{\mathrm{s}} 
+ \frac{2t_{\mathrm{b}}^2}{\Delta\varepsilon},
\\
\varepsilon_{\mathrm{g}}
&=&
\varepsilon_0 + t_{\mathrm{s}} ,
\\
\varepsilon_{\mathrm{u'}}
&\approx&
\varepsilon_0 - \Delta\varepsilon
- \frac{2t_{\mathrm{b}}^2}{\Delta\varepsilon}.
\end{eqnarray}%
\label{eq:eigenenergy}%
\end{subequations}
The respective wave functions are given by
\begin{subequations}
\begin{eqnarray}
\varphi_{\mathrm{u}} 
&=&  
\frac{1}{\sqrt{2+a^2}}
[(\varphi_{\mathrm{L}}+\varphi_{\mathrm{R}})-a \varphi_{\mathrm{C}}],
\label{eq:wf-u}
\\
\varphi_{\mathrm{g}} 
&=& 
\frac{1}{\sqrt{2}}(\varphi_{\mathrm{L}}-\varphi_{\mathrm{R}}),
\\
\varphi_{\mathrm{u'}} 
&=&
\frac{-1}{\sqrt{1+a^2/2}}
\left[\varphi_{\mathrm{C}} 
 + \frac{a}{2}
  (\varphi_{\mathrm{L}}+\varphi_{\mathrm{R}})
\right],
\label{eq:wf-u'}
\end{eqnarray}%
\label{eq:wf}%
\end{subequations}
where the quantity $a$ is
given by $a\approx (2t_\mathrm{b}/\Delta \varepsilon)$. 
In this picture, the L, R and C fragments wave functions 
are assumed to be orthogonal which is
obviously not a limitation for the description.
By comparing Eq.\ (\ref{eq:wf}) with the u and g MOs shown in Fig.\
\ref{fig:mo-level}, one can conclude distinctly that the wave functions
$\varphi_{\mathrm{u}}$ and $\varphi_{\mathrm{g}}$ correspond to the u and g
MOs, respectively.
To scope out the $\varphi_{\mathrm{u'}}$ MO which is given in
Eq.\ (\ref{eq:wf-u'}), we look into the set of ungerade MOs 
obtained from the SCF calculations.
The in-phase combination u$_\mathrm{inner}$ of $\varphi_{\mathrm{L}}$,
$\varphi_{\mathrm{R}}$, and $\varphi_{\mathrm{C}}$ wave functions is shown in
Fig.\ref{fig:energydiagram-mo} and its energy is
$\varepsilon_{\mathrm{u_{inner}}}=-19.59$~eV. Since a typical value for
$|t_{\mathrm{b}}|$ is of the order of several eV, 
the latter energy separation fully
justifies the approximate expressions given previously.  Thus, we can assign
the $\varphi_{\mathrm{u'}}$ wave function to the u$_\mathrm{inner}$ MO
identified in the SCF calculations.

\begin{figure}[t]
\includegraphics[width=6cm]{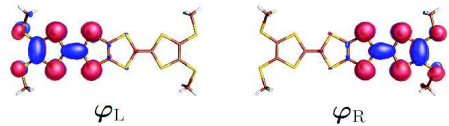}
\caption{
The L and R MOs obtained through a unitary transformation 
of Eq.\ (\ref{eq:wf}) by taking $a=0.25$.
}
\label{fig:TTM-TTP_MO_LR}
\end{figure}

The ratio $2t_\mathrm{b}/\Delta\varepsilon(\approx a)$ was estimated
from the charge distribution in the [TTM-TTP]$^+$ ion, i.e.,
so-called Mulliken charges.  Based upon the CAS[3,2] wave functions,
CASPT2 calculations were performed and the resulting Mulliken charges
were estimated.  Following the three-fragment model, we summed up the
Mulliken charges within each fragment, {\it e.g.},
$\rho_\mathrm{L}=\sum_{i\in \mathrm{L}} \rho_i$ 
where $\rho_i$ is the charge of the respective atoms.
In the ground state configuration where the u MO is singly
occupied, the Mulliken charges on the respective fragments of the
[TTM-TTP]$^{+}$ ion were estimated as
\begin{eqnarray}
\rho_\mathrm{L}= \rho_\mathrm{R} = +0.634,
\quad
\rho_\mathrm{C} = -0.268 .
\end{eqnarray}
The total charge is $\rho_\mathrm{L}+\rho_\mathrm{R}+\rho_\mathrm{C}
 =+1$. 
 In order to extract the contribution from the SOMO,
we analyzed the charge difference between the cation [TTM-TTP]$^{+}$
and the [TTM-TTP] molecule.  The latter exhibits the
following Mulliken charges on the respective fragments
 $\rho_\mathrm{L}^0=
\rho_\mathrm{R}^0 = + 0.149$ and $\rho_\mathrm{C}^0 = -0.297$. 
 Thus, the
charge differences $\Delta \rho_Z \equiv (\rho_Z-\rho_Z^0)$
($Z=\mathrm{L,R,C}$) were calculated as
\begin{eqnarray}
\Delta \rho_\mathrm{L}= \Delta \rho_\mathrm{R} = 0.485,
\quad
\Delta \rho_\mathrm{C} = 0.029.
\label{eq:Deltarho_1mol}
\end{eqnarray}
This means that the hole is mainly localized on the L and R fragments,
whereas the charge on the C fragment remains almost unchanged.  By
combining the information upon the SOMO [Eq.\ (\ref{eq:wf-u})] and the
numerical data of the charge distribution [Eq.\
(\ref{eq:Deltarho_1mol})], we can estimate the parameter $a\approx
(2t_\mathrm{b}/\Delta \varepsilon)$ by using the relations
$1/(2+a^2)=\Delta \rho_\mathrm{L}$ or $a^2/(2+a^2)=\Delta
\rho_\mathrm{C}$.  Based on this analysis, we find $a\approx 0.25$ and
finally the ratio $(t_\mathrm{b}/\Delta \varepsilon)$ is $\approx 0.12$.
Once the parameter $a$ is determined, the fragment wave functions can be
obtained explicitly through the inverse unitary transformation of Eq.\
(\ref{eq:wf}).  The resulting L and R MOs are shown in Fig.\
\ref{fig:TTM-TTP_MO_LR} and correspond to the $\varphi_{\mathrm{L}}$ and
$\varphi_{\mathrm{R}}$ local MOs.  Since
$[(\varepsilon_\mathrm{u}+\varepsilon_\mathrm{g})/2 -
\varepsilon_\mathrm{u_{inner}}] = \Delta \varepsilon [1 + 3
(t_\mathrm{b}/\Delta\varepsilon)^2]$ [see Eq.\ (\ref{eq:eigenenergy})],
the ROHF eigenvalues $\varepsilon_\mathrm{u}=-8.27$ eV,
$\varepsilon_\mathrm{g}=-8.69$ eV, and
$\varepsilon_\mathrm{u_{inner}}=-19.59$~eV lead to the effective energy
of the C fragment $\Delta\varepsilon \approx 10.6$ eV.  Finally, the
magnitudes of the through-bond/through-space interactions can be
estimated, $t_\mathrm{b} \approx 1.32$ eV and $t_\mathrm{s} \approx
-0.04$ eV.  From our evaluation, $t_\mathrm{s}$ is almost negligible,
and the bonding and antibonding MOs of Fig. \ref{fig:energydiagram-mo}
are almost degenerated.  The through-space integral is much smaller than
the through-bond one, $|t_{\mathrm{b}}/t_{\mathrm{s}}|\approx 28$.
Since the energy difference between the u and g MOs is determined from
Eq.\ (\ref{eq:eigenenergy})
\begin{eqnarray}
\varepsilon_{\mathrm{u}}
-
\varepsilon_{\mathrm{g}}
\approx
 - 2t_{\mathrm{s}} 
+ \frac{2t_{\mathrm{b}}^2}{\Delta\varepsilon},
\end{eqnarray}
the origin of the quasidegeneracy of the u and g MOs can be clarified
in the light of the extracted parameters.  

In conclusion, the u and g MOs ordering is completely determined by the
relative energetics of this three-piece molecule.
The through-space interaction $t_\mathrm{s}$ is almost negligible,
while through-bond interaction 
$t_\mathrm{b}$ determines the energy spectrum.

\section{Intramolecular charge ordering}

In this section, the interactions between neighboring 
 [TTM-TTP]$^{+}$ cations have been investigated
using a dimer extracted  from 
  the (TTM-TTP)I$_3$ crystal. 
We used the atomic coordinates of the crystal structure at 298 K.
If the crystal-structure data at low-temperature phase were available,
 we may perform 
  more quantitative analysis by combining the geometrical optimization.
However, it can be considered that the following results are not
 affected qualitatively.
Since we were not only interested in charge distribution,
CASPT2 calculations were also performed
to specify the low-energy spectroscopy
of the [TTM-TTP]$_2^{2+}$ dimer.

\subsection{``MOs''}

\begin{figure}[t]
\includegraphics[width=6cm]{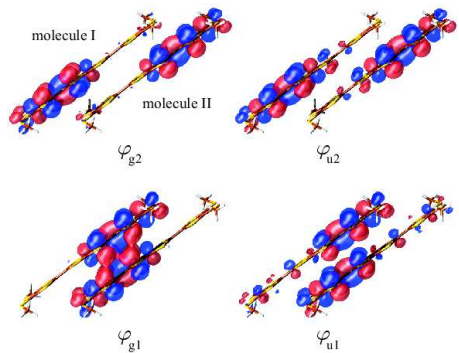} 
\caption{
Valence MOs of the [TTM-TTP]$_2^{2+}$ system
obtained from CAS[6,4]SCF \textit{ab initio}
 calculations upon the singlet state of g symmetry. 
}
\label{fig:mo2}
\end{figure}

First, CASSCF calculations were carried out on the [TTM-TTP]$_2^{2+}$ dimer.
These calculations were performed including six electrons in four MOs in the
active space (CAS[6,4]) to account for the important static correlation
effects.  
The interactions between the neighboring [TTM-TTP]$^{+}$ ions labeled as
\textbf{I} and \textbf{II}
 give rise to the effective MOs
shown in Fig.\ \ref{fig:mo2}.  Importantly, the dimer exhibits an inversion
center and the resulting MOs can be classified into gerade and
ungerade.  Nevertheless, the inversion center of each individual
subunit is lost, which might lead to electron localization within the
[TTM-TTP]$^{+}$ building blocks. 
As expected, the CAS[6,4]SCF frontier orbitals (Fig.\ \ref{fig:mo2})
 consist in the in-phase and
out-of-phase combinations of the
 $\varphi_{\mathrm{L},\I}$, 
 $\varphi_{\mathrm{R},\I}$,
 $\varphi_{\mathrm{L},\II}$, and
 $\varphi_{\mathrm{R},\II}$ MOs.
The  $\varphi_{\mathrm{L},\I}$ and
 $\varphi_{\mathrm{R},\I}$ are the left (L) and right (R)  
 localized MOs on the subunit \textbf{I}
(see Fig.\ \ref{fig:TTM-TTP_MO_LR}). 
A similar definition holds for the subunit \textbf{II}.

From the subsequent CASPT2 treatment, the ground state is a gerade
singlet, and the active MOs exhibit occupation numbers $2.0$, $2.0$,
 $0.9$, and $1.1$, respectively.
Importantly, the occupation numbers of the g$_2$ and u$_2$ MOs strongly
deviate from 2 and 0, a feature of the open-shell nature of the ground state
singlet.  The symmetry breaking within each [TTM-TTP]$^{+}$ ion
induced by the presence of a second partner leads to a mixing of the
orbitals depicted in Fig.\ \ref{fig:energydiagram-mo}.
This particular mechanism is likely to result in the charge ordering 
we now wish to examine.

\subsection{Mulliken charge}

From the CASPT2 calculations
performed upon the [TTM-TTP]$_2^{2+}$ dimer, a
similar Mulliken charge analysis was carried out to quantify the charge
redistribution accompanying the dimer formation.
The ground singlet state  exhibits the following charge reorganization
as compared to the isolated neutral [TTM-TTP] molecule
\begin{eqnarray}
\Delta\rho_\mathrm{L} = 0.697 , \quad
\Delta\rho_\mathrm{R} = 0.272 , \quad
\Delta\rho_\mathrm{C} = 0.030 .
\end{eqnarray}
As a major conclusion, the charge on the left (right) fragment is strongly
enhanced (reduced), which is a clear indication of the intramolecular charge
ordering.  Again, the charge on the center fragment C is almost unchanged.
The present evaluation based on the two-molecule dimer overestimates the charge
difference between the L and R fragments as compared to its value in the
crystal, and also the validity of Mulliken charges remains questionable since
the fluctuation effects were not included.  
Nevertheless, this qualitative analysis supports the intramolecular
CO state.

\subsection{Low-energy spectroscopy}

Our Mulliken charge analysis is suggestive of an
intramolecular charge ordering mechanism in the (TTM-TTP)I$_3$
material. Thus, one may expect a charge localization on the left or
right parts of the TTM-TTP building blocks. Such scenario is likely to
give rise to magnetic interactions involving either the inner parts
(i.e., $\varphi_{\mathrm{R},\I}$ and $\varphi_{\mathrm{L},\II}$
fragments MOs) or the outer parts (i.e., $\varphi_{\mathrm{L},\I}$
and $\varphi_{\mathrm{R},\II}$).

\begin{figure}[b]
\includegraphics[width=7cm]{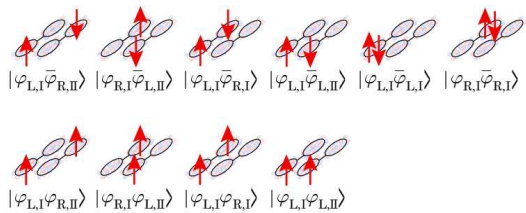}
\caption{
Schematic view of the different configurations  
using a hole picture for the singlet (top) and
triplet (bottom), respectively.
The subscripts $\I$ and $\II$ stand for the  molecules $\I$ and $\II$
 as depicted in Fig.\ \ref{fig:mo2}. 
}
\label{fig:orderedstate}
\end{figure}

\begin{table*}[bth]
\caption{
Wave functions decompositions (weights) of the
[TTM-TTP]$_2^{2+}$ low-energy states. The wave functions are expressed in
terms of the two hole determinants. CAS[6,4]PT2 energies with respect to
the ground state singlet g are given in eV.  }
\label{table:OccupationNumber}
\begin{tabular}{lcccccccc}
\hline\hline
& \hspace*{.1cm} $|\,\varphi_{\mathrm{L},\I} \, \bar\varphi_{\mathrm{R},\II}\rangle$ \hspace*{.1cm}  &  \hspace*{.1cm} 
   $|\,\varphi_{\mathrm{R},\I} \, \bar\varphi_{\mathrm{L},\II}\rangle$ \hspace*{.1cm}  &  \hspace*{.1cm} 
   $|\,\varphi_{\mathrm{L},\I} \, \bar\varphi_{\mathrm{R},\I}\rangle$ \hspace*{.1cm}  &  \hspace*{.1cm} 
   $|\,\varphi_{\mathrm{L},\I} \, \bar\varphi_{\mathrm{L},\II}\rangle$\hspace*{.1cm}   &  \hspace*{.1cm} 
   $|\,\varphi_{\mathrm{L},\I} \, \bar\varphi_{\mathrm{L},\I}\rangle$ \hspace*{.1cm}  &  \hspace*{.1cm} 
   $|\,\varphi_{\mathrm{R},\I} \, \bar\varphi_{\mathrm{R},\I}\rangle$  
\hspace*{.3cm} && Energy
\\
& ($\%$)& ($\%$)& ($\%$)& ($\%$)& ($\%$)& ($\%$) && (eV)
\\  
Singlet, gerade   
& $\bm{84}$ & $1$ & $3$ & $9$  & $3$  & $0$  && $E_0=0.00$
\\
& $11$  & $27$  & $\bm{44}$  & $13$  & $5$  & $0$  && $E_2=0.27$ 
\\ 
\hline
\end{tabular} 
\vspace*{.5cm}
\begin{tabular}{lcccccc} 
& \hspace*{.5cm}
   $|\,\varphi_{\mathrm{L},\I} \, \varphi_{\mathrm{R},\II}\rangle$ \hspace*{.5cm} &  \hspace*{.5cm}
   $|\,\varphi_{\mathrm{R},\I} \, \varphi_{\mathrm{L},\II}\rangle$ \hspace*{.5cm} &  \hspace*{.5cm}
   $|\,\varphi_{\mathrm{L},\I} \, \varphi_{\mathrm{R},\I}\rangle$  \hspace*{.5cm}&  \hspace*{.5cm}
   $|\,\varphi_{\mathrm{L},\I} \, \varphi_{\mathrm{L},\II}\rangle$ \hspace*{.5cm} &&  
   Energy 
\\
& ($\%$)& ($\%$)& ($\%$)& ($\%$) && (eV)
\\   
Triplet, ungerade 
\hspace*{.5cm}
& $\bm{62}$  & $5$  & $0$  & $33$  && $E_1=0.09$
\\
& $28$  & $\bm{47}$  & $2$  & $23$  && $E_3=0.47$
\\
& $6$  & $8$  & $\bm{70}$  & $16$  && $E_4=0.62$
\\
\hline\hline
\end{tabular} 
\end{table*}

Thus, starting from the CAS[6,4]SCF calculations, the low-energy
spectroscopy of the [TTM-TTP]$_2^{2+}$ species was inspected. 
The multi-reference CASSCF wave functions
were expanded using a local orbitals basis set  
$\{ 
\varphi_{\mathrm{L}, \I}$, 
$\varphi_{\mathrm{R},\I}$,
$\varphi_{\mathrm{L},\II}$,
$\varphi_{\mathrm{R},\II} \}$ 
following the transformation
\begin{subequations}
\begin{eqnarray}
\varphi_{\mathrm{g1}}
&\approx& 
\frac{1}{\sqrt{2}} 
(
- \varphi_{\mathrm R,\I}
+ \varphi_{\mathrm L,\II}
),
\\
\varphi_{\mathrm{u1}}
&\approx&
\frac{1}{\sqrt{2}}
(
- \varphi_{\mathrm R,\I}
- \varphi_{\mathrm L,\II}
),
\\
\varphi_{\mathrm{g2}}
&\approx& 
\frac{1}{\sqrt{2}}
(
+ \varphi_{\mathrm L,\I}  
- \varphi_{\mathrm R,\II}
),
\\
\varphi_{\mathrm{u2}}
&\approx& 
\frac{1}{\sqrt{2}}
(
- \varphi_{\mathrm L,\I}  
- \varphi_{\mathrm R,\II}
).
\end{eqnarray}
\end{subequations}
Since we are dealing with a six-electron/four-MO system,
 we performed our analysis in the two-hole/four-MO picture.
Schematic representations of the different hole
configurations for the singlet and triplet states based on these local
orbitals are shown in Fig.\ \ref{fig:orderedstate}.
 Such transformation
affords a reading of the different wave functions and the extractions of the
relevant information in a valence-bond type analysis.  Since $|t_s|$ is
relatively small, the left and right moieties within each unit do not
significantly overlap. Thus, the previous transformation is almost
unitary.

The lowest lying states of the [TTM-TTP]$_2^{2+}$ species are given in
Table \ref{table:OccupationNumber}. 
The ground state is singlet g and consistent with a picture that minimizes 
 the electrostatic energy between the two [TTM-TTP]$^+$  ions.
The comparison between the ground
state singlet g and first triplet u is very instructive. In both cases,
the wave function is largely dominated by the electronic configurations 
(84 \% and 62 \%
in Table \ref{table:OccupationNumber} for the singlet and triplet
states, respectively) involving the inner parts (see Fig.\
\ref{fig:orderedstate}). 
The energy difference affords an
evaluation of the intermolecular magnetic exchange interaction :
\begin{equation}
J \equiv E_1-E_0
\approx 0.09 \,\, \mbox{eV}
\label{eq:J}
\end{equation}
A positive value reflects an antiferromagnetic interaction, resulting
from the intramolecular charge ordering. 
Such a value suggests that the spin gap excitation energy should be 
  $\approx 900$ K.
 The first-excited singlet
corresponds to the holes localization within one [TTM-TTP]$^+$ ion. 
This is a reflection of an \textit{intermolecular} charge
ordering which lies much higher and might not be relevant to describe
the low-energy properties of the (TTM-TTP)I$_3$ material.  The energy
difference (0.35 eV) between this second excited state (singlet g) and
the fourth one (triplet u) supports the strong antiferromagnetic
interaction within each subunit. This is a mechanism that involves again
intermolecular electron transfer.  
The low-energy spectroscopy in the (TTM-TTP)I$_3$ material
should be mostly
governed by the through-space interdimer magnetic interaction
$J$, which results from the intramolecular charge ordering.

\section{Summary and Discussions}

In the present paper, we examined the low-energy properties of the
(TTM-TTP)I$_3$ material.  Complementary semi-empirical and
wave function-based \textit{ab initio} calculations were performed upon
the elementary unit and a dimer to investigate the underlying electronic
distribution.  We showed that the SOMO and HOMO-1 are close in energy
and that a one-band picture should be ruled out.  A chemical
understanding arises from the inspection of the building block and
effective parameters extracted from a three-fragment model allowed us to
rationalize the quasidegeneracy of the [TTM-TTP] frontier MOs.  
From the calculation of a system consisting of two neighboring
molecules, the inversion symmetry within each molecule is lost. This
result is in agreement with experimental data. Indeed, the
intramolecular charge ordering which has been experimentally suggested
actually occurs in the two-molecule cluster system. The mixing of the
SOMO and HOMO-1 results in an electron trapping which results in a 
 0.697 / 0.272 on the L / R moieties of the [TTM-TTP]$^+$ ion.  
Finally, our {\it ab initio} calculations suggest
that the low-energy of the (TTM-TTP)I$_3$ material is controlled by
a single exchange interaction $0.09$ eV, resulting from the charge ordering.

In order to complement our Mulliken charge analysis, we also performed
preliminary Raman and infrared (IR) calculations upon the [TTM-TTP]$^+$ unit.
It has been observed upon cooling that the Raman 1490~cm$^{-1}$ band splits
into two peaks centered at 1487~cm$^{-1}$ and 1499~cm$^{-1}$, where this
phenomenon was attributed to the differentiation between the C$=$C ylidene
bonds, featuring a symmetry breaking.  \cite{Yakushi2003} We used the
{\sc gaussian03} package \cite{Gaussian} on isolated one- and two-molecule systems,
and full geometry optimizations were carried out on both systems.
The information extracted from the one-molecule system should be compared with
the high-temperature regime. In contrast, the two-molecule system is expected
to give access to the low-temperature Raman spectrum characteristics.  The
calculated spectrum of two-molecule system displays two vibrational frequencies,
1522 cm$^{-1}$ and 1534 cm$^{-1}$, while that of the 1-molecule system shows a
single band at 1535 cm$^{-1}$.  Despite a general blue-shift, this result
is in agreement with experimental data, and can be attributed to an
intramolecular CO phase accompanying the descent in symmetry within the
TTM-TTP units.  In this calculation, the [TTM-TTP]$_2^{2+}$ dimer was
extracted from the available crystallographic data.  A direct comparison with
experimental findings would rely on Raman and IR calculations using the
low-temperature crystal structure. Unfortunately, the lack of such x-ray data
disposes of this strategy.

Throughout this work, we have focused only on the [TTM-TTP] ion
and have neglected the effect of the counteranion of I$_3^-$.  As a matter
of fact, it has been pointed out, from X-ray
measurement,\cite{Fujimura1999} that the large displacement of the I$_3$
species occurs in the low-temperature phase.  This might yield changes
in the electrostatic potential, and possibly the electronic distribution in
the [TTM-TTP] ion would qualitatively be modified.  This point
is to be studied in future work.

\vspace*{.2cm}
\noindent
\textbf{Acknowledgements}

MT thanks T.\ Kawamoto, T.\ Mori, H.\ Kobayashi, and  S.\ Yasuzuka,
  for the fruitful discussions on the (TTM-TTP)I$_3$ compounds.
MLB, MT, YO, and YS thank S.\ Ishibashi and H.\ Seo for 
  the discussions on the theoretical aspects.
MT thanks S.\ Yasuzuka for the enlightening discussions
  in the early stage of the present work.
MLB was supported by the JSPS postdoctoral fellowship for Foreign Researchers.
This research was partially supported by
Nagoya University Science Foundation and
Grant-in-Aid for Scientific Research on Innovative Areas 
(20110002)
from the 
Ministry of Education, Culture, Sports, Science and Technology, Japan.

\end{document}